\documentclass[conference]{IEEEtran}
\IEEEoverridecommandlockouts
\usepackage{cite}
\usepackage{amsmath,amssymb,amsfonts}
\usepackage{graphicx}
\usepackage{textcomp}
\usepackage{float}
\begin{document}

\title{Test Automation with Grad-CAM Heatmaps -- A Future Pipe Segment in MLOps for Vision AI?}

\author{\IEEEauthorblockN{Markus Borg\IEEEauthorrefmark{1}\IEEEauthorrefmark{2},
Ronald Jabangwe\IEEEauthorrefmark{1},
Simon Åberg\IEEEauthorrefmark{2},
Arvid Ekblom\IEEEauthorrefmark{2}, 
Ludwig Hedlund\IEEEauthorrefmark{2} and
August Lidfeldt\IEEEauthorrefmark{2}}
\IEEEauthorblockA{\IEEEauthorrefmark{1}RISE Research Institutes of Sweden, \{markus.borg, ronald.jabangwe\}@ri.se}
\IEEEauthorblockA{\IEEEauthorrefmark{2}Dept. of Computer Science, Lund University, Lund, Sweden, \{si6365ab-s, ar7180ek-s, lu3746he-s, au4322li-s\}@student.lu.se}}

\maketitle
\thispagestyle{plain}
\pagestyle{plain}

\begin{abstract}
Machine Learning (ML) is a fundamental part of modern perception systems. In the last decade, the performance of computer vision using trained deep neural networks has outperformed previous approaches based on careful feature engineering. However, the opaqueness of large ML models is a substantial impediment for critical applications such as in the automotive context. As a remedy, Gradient-weighted Class Activation Mapping (Grad-CAM) has been proposed to provide visual explanations of model internals. In this paper, we demonstrate how Grad-CAM heatmaps can be used to increase the explainability of an image recognition model trained for a pedestrian underpass. We argue how the heatmaps support compliance to the EU's seven key requirements for Trustworthy AI. Finally, we propose adding automated heatmap analysis as a pipe segment in an MLOps pipeline. We believe that such a building block can be used to automatically detect if a trained ML-model is activated based on invalid pixels in test images, suggesting biased models.
\end{abstract}


\section{Introduction}
Machine Learning (ML) is increasingly used in Cyber-Physical Systems (CPS). In autonomous vehicles, ML using Deep Neural Networks (DNN) enables considerably more accurate perception models compared to conventional hand-crafted computer vision techniques. However, as the perception models will increasingly be used as components in safety-critical vehicular functions~\cite{falcini_deep_2017}, they must be duly tested. ML-based perception must constitute ``Trustworthy AI,'' defined as lawful, ethical, and robust by the High-Level Expert Group on AI (AI-HLEG) appointed by the European Union~\cite{high-level_expert_group_on_artificial_intelligence_ethics_2019}. Especially the robustness quality is an appropriate target for test automation, but also legal and ethical compliance could be tested.  

A well-known issue with DNNs is the limited explainability of their output~\cite{adadi_peeking_2018,shen2020interpretability}. In a DNN with a number of trained weights in the magnitude of $10^7$, it is difficult to explain why a specific output has been correctly perceived by the model. ML-based solutions are often regarded as black boxes whose internals are hard to test. Furthermore, ML-models are unfortunately known to be sensitive to minor image perturbations~\cite{azulay_why_2019}. To address these issues, Selvaraju \textit{et al.} proposed visual explanations using Gradient-weighted Class Activation Mapping (Grad-CAM)~\cite{selvaraju2017grad-cam}. Grad-CAM identifies areas with high neuron activation in a DNN's last convolutional layer and visualizes the results as an activation heatmap superimposed over an input image. Several authors have reported successful applications of using Grad-CAM to explain the inner workings of DNNs. Examples include iris classification~\cite{Trokielewicz2018}, legal text processing~\cite{gorski2020towards}, and Android malware detection~\cite{9307643}. In this paper, we propose using Grad-CAM as part of test automation.

ML development is highly iterative with considerable experimentation to identify the most promising models. Data version control and experiment tracking are essential components to succeed, as well as sophisticated automation infrastructures. Experience from industrial-grade ML projects has stressed the importance of MLOps pipelines that automate steps from data collection to model deployment. For any proposed ML testing approach to be industrially relevant, they must be integrable in contemporary MLOps pipelines. We posit that relevant ML testing research shall be integrated into an MLOps pipeline that interweaves source code and data~\cite{borg_aiq_2021}.

The discussion in this paper, the starting point of a new research direction, is guided by two research questions.
\begin{itemize}
\item[RQ1] How can Grad-CAM heatmaps support development of trustworthy ML-based perception?
\item[RQ2] How can analysis using Grad-CAM heatmaps support development in an MLOps context? 
\end{itemize}

To explore these questions, we present a proof-of-concept that generates Grad-CAM heatmaps for a common convolutional neural network architecture for object recognition. Based on our previous work on object recognition in an underpass in the city of Helsingborg~\cite{lidfeldt2020enabling}, we demonstrate how Grad-CAM heatmaps provide actionable insights to developers. As an example, if the ML model classifies an image as containing a bicyclist, pixels close to the bicycle shall be decisive to the neural network -- relying on other pixels would be invalid, suggesting a biased ML model. We describe how we use Grad-CAM heatmaps to manually analyze neuron activations corresponding to both correct and incorrect classifications. Finally, we present how we plan to extend this work from automatic generation of Grad-CAM heatmaps to also include automatic analysis of test results, by comparing the position of activated pixels to ground truth bounding boxes. Our goal is to implement this approach in an MLOps pipeline for an ongoing development project of a ML-based advanced driver-assistance system. All source code and some non-sensitive image data are available on GitHub under an open source software license.

%Our vision involves automatic generation and analysis of Grad-CAM heatmaps as explicit steps in an MLOps pipeline. For object detection, it is important to verify that the DNN considered the right pixels to make predictions. We envision that based on an annotated dataset, i.e., ground truth bounding boxes around objects, Grad-CAM heatmaps could be used to automatically test that a DNN was activated by valid parts of an image. 

The rest of the paper is organized as follows. Section~\ref{sec:bg} introduces 1) convolutional neural networks and Grad-CAM, 2) the emerging topic of ML testing, and 3) AI-HLEG's definition of trustworthy AI. In Section~\ref{sec:case}, we briefly describe the application under study. Section~\ref{sec:poc} presents our proof-of-concept and several Grad-CAM heatmaps. In Section~\ref{sec:disc}, we discuss our proof-of-concept in the light of the research questions. Finally, Section~\ref{sec:conc} concludes the paper.

\section{Background and Related Work} \label{sec:bg}

\subsection{Convolutional Neural Networks and Grad-CAM}
An artificial neural network (ANN) is one out of many ML models that can find complex patterns in large datasets. The concept builds on layers of weighted nodes connected by edges. When an input is provided to the network, signals propagate through the structure with varying strengths depending on the node weights. In the last decade, the amount of available data and compute increased substantially. Researchers found that stacking more and more layers of neurons resulted in better and better output, i.e., deep neural networks could harvest the potential in vastly available data and massively parallel hardware architectures could fit the models in reasonable time.

Computer vision is one application area that experienced significant breakthroughs using deep neural networks. The sequential model structure often used in image recognition is composed of an input layer followed by a certain number of connected layers, in turn leading to an output layer. The model input is usually an image and each node in the input layer is fed the value of a certain pixel from that image. The output of the network is usually a prediction of the contents of the image, e.g., image recognition (categorizing the content) and image detection (localizing objects).

Convolutional layers are key building blocks in deep neural networks for computer vision, referred to as Convolutional Neural Networks (CNN). In a CNN, convolutional layers act as filters to an input keeping important visual features. The filter, smaller than the entire image, is gradually moved over the entire input space to produce a feature map. The feature map indicates where features were detected, and how strong the activations for that feature were.

In CNNs, it is hard to analyze what individual nodes represent in the images and how they contribute to the final predictions. To support the explainability, Selvaraju \textit{et al.} developed Grad-CAM~\cite{selvaraju2017grad-cam}. Given a specific class label and input, Grad-CAM visualizes areas with high neuron activation in the last convolution layer of a CNN. The activation is then presented in the form of an activation heatmap superimposed on the input image. Using this approach, a human can assess whether the areas of the image that were the most important in the classification make sense. As an example, Selvaraju \textit{et al.} presents how Grad-CAM can be used to identify how classifications are influenced by biased training data. The authors show how a nurse-or-doctor image classifier primarily activated on gender cues such as hairstyle rather than equipment or job tasks. We return to the discussion on bias in Section~\ref{sec:trust}.

%The class localization activation maps \(L_{Grad-CAM}^c\) is calculated by:
%\begin{equation}
%L_{Grad-CAM}^c = ReLu(\sum_k\alpha_k^c A^k)\
%\end{equation}
%Where \(\alpha_k^c\) are the neuron weights which are calculated as:
%\begin{equation}
%\alpha_k^c = \frac{1}{Z}\sum_i\sum_j \frac{\partial y^c}{\partial A_ij^k}
%\end{equation}
%\(\frac{\partial y^c}{\partial A_ij^k}\) is the gradients via backpropagation and \(\frac{1}{Z}\sum_i\sum_j\) is the global average pooling. The rectified linear activation function, shortened as ReLu, nullifies subzero values and is applied to only extract the neurons that have a positive effect on the class of interest.\cite{8237336}

\subsection{Machine Learning Testing}
Systems that rely on ML brings entirely new challenges to the field of software testing. As opposed to traditional software, a large part of the functionality in ML software is trained rather than coded. The implications of this is that the model might do exactly what it was trained for but still produce erroneous behaviour. This behaviour can be caused by many different factors such as errors in the data collection, data processing, model selection or model training. If any of these steps are performed improperly biases or incorrect examples may be infused in the underlying dataset and thus into the model training~\cite{marijan2019challenges}.

In traditional software testing, the written tests are there to ensure that the written logic of the software under test follows the expectations of the designer or developer. While we are interested in the result, the way that the software arrives at the result is of equal importance. A tester has tools such as unit tests and code coverage to his at her disposal, to examine atomic pieces of the software to ensure that they behave as expected, and to ensure that all relevant parts of the software have been thoroughly analyzed. Comparing ML testing to software testing, the inner workings of the ML model are often opaque, disqualifying traditional testing tools. Several authors have highlighted this challenge in safety-critical automotive software~\cite{salay2018using,borg2019safely}, and novel safety standards such as ISO/PAS 21448 Safety of the Intended Functionality (SOTIF) are evolving to complement the previous source code-based logic~\cite{international_organization_for_standardization_isopas_2019}. Currently, ML testing relies primarily on performance metrics to determine the quality of a model \cite{marijan2019challenges}. 

Two primary factors encumber ML testing. First, the complexity of the ML models has increased substantially. State-of-the-art image recognition architectures contain hundreds of millions of weights~\cite{hao2019training}. Second, the size of the training datasets. To fit the complex ML models, data scientists train them on enormous datasets. Today the sheer size of many datasets makes getting a complete understanding of the data characteristics and their implications beyond human comprehension. Tesla presents their autopilot as an example of a cutting-edge ML-based automotive system. Tesla Autopilot contains 48 different neural networks, making 1,000 predictions per second. These models have been trained for a combined total of 70,000 GPU hours~ \cite{tesla}. While Tesla Autopilot is definitely one of the most complex ones on the market, the challenge of maintaining testability and interpretability generalizes down to systems that rely on less complex ML models. ML testing requires novel approaches.

Testing ML-based perception models goes beyond reporting standard metrics for classification accuracy. Several researchers have proposed approaches to test ML applications that process image data~\cite{zhang_machine_2020}. Examples include, measuring the proportion of neurons covered by a certain test set, analogous to code coverage testing of software~\cite{pei_deepxplore_2019}, estimating how different input is compared to what it learned during training (surprise adequacy)~\cite{kim_guiding_2019} to guide future training efforts, and displaying adversarial inputs on billboards in order to test robustness of the autonomous driving systems~\cite{patel2019adaptive}. 

\subsection{Trustworthy AI} \label{sec:trust}
AI-HLEG states that Trustworthy AI must be: 1) Lawful – complying with all applicable laws and regulations; 2) Ethical – ensuring adherence to ethical principles and values; 3) Robust – both from a technical and social perspective~\cite{high-level_expert_group_on_artificial_intelligence_ethics_2019}. AI-HLEG further specifies seven ``key requirements'' that Trustworthy AI systems must adhere to:
\begin{itemize}
\item[R1] \textbf{Human agency and oversight}. An AI system shall ``empower human beings, allowing them to make in-formed decisions and fostering their fundamental rights.'' Oversight mechanisms must allow humans to inspect the decisions made.
\item[R2] \textbf{Technical robustness and safety}. An AI system shall be ``resilient and secure.'' Prevention of harm must permeate the design of the system. Security is fundamental to protect against antagonistic attacks. Safety is essential to ensure operation without harming people or the environment. Furthermore, AI systems must be sufficiently accurate and reliable.
\item[R3] \textbf{Privacy and data governance}. An AI system shall ``ensure full respect for privacy and data protection.'' This requires adequate data management policies and data governance mechanisms to assure the quality and integrity of the data. 
\item[R4] \textbf{Transparency}. The AI system shall be open about the underlying ``data, system, and AI business models.'' Output must be explained, and users shall be informed of inherent capabilities and limitations.
\item[R5] \textbf{Diversity, non-discrimination and fairness}. An AI system shall be ``accessible to all, regardless of any disability, and involve relevant stakeholders throughout their entire life circle.'' Unfair bias can have severe negative implications and must be avoided.
\item[R6] \textbf{Societal and environmental well-being}. An AI system shall ``benefit all human beings, including future generations.'' This requirement covers how AI affects the environment and society at large.
\item[R7] \textbf{Accountability}. An AI system shall ``ensure responsibility and accountability for AI systems and their outcomes.'' Mechanisms must be in place to enable auditability, i.e., the assessment of algorithms, data and design processes. 
\end{itemize}

To support development organizations with actionable advice, AI-HLEG published the Assessment List for Trustworthy Artificial Intelligence (ALTAI)~\cite{high-level_expert_group_on_artificial_intelligence_assessment_2020}. ALTAI was developed as a checklist intended for self-assessment of compliance to the seven key requirements [R1-R7]. We argue that automated testing with Grad-CAM in an MLOps pipeline could support demonstration of ALTAI compliance, especially when evolving ML models until they are production-ready. More specifically, as we discuss in Section~\ref{sec:poc}, we believe that our solution proposal can to various extent support R1, R2, R4, R5, and R7.

%If we are to allow machine learning systems to handle an increasingly important decision making in our society it is integral that we can build trust into these systems. This trust depends on reassuring that the systems perform both correctly and fairly. To facilitate the technology adoption standard testing practices should be adopted. Preferably these should be made easily accessible and automated to limit manual testing and configuration, as time is often precious when machine learning models have to be trained for hours to weeks. 

%With machine learning models becoming more prevalent every year with the industry continuously finding new applications, the need for testing machine learning models increases. An important factor in machine learning testing is the interpretability of the model to increase trust for the system. When machine learning is applied to high-risk situations such as autonomous vehicles where human lives are at stake, being able to verify that the models follow the intended logic of the designer becomes even more important. \cite{zhang_machine_2020}

\section{Case Description} \label{sec:case}
The case under study is a motion-activated network camera mounted in an underpass located in Helsingborg, Sweden. The camera was initially installed as part of a research project on crime prevention using Internet-of-Things technology, i.e., to enable early detection of illegal graffiti -- whose removal costs are substantial for municipalities. We have the permission to use collected non-sensitive data in other research projects related to computer vision.

In a previous study~\cite{lidfeldt2020enabling}, we used images from the underpass to train various DNNs for image recognition. Ninety minutes of high resolution video clips were collected over five days in 2020. We split the video clips into individual frames at a frame rate of one per second resulting in a dataset containing. Inspired by automotive engineering, we specify the Operational Design Domain (ODD)~\cite{gyllenhammar2020towards} of our classification model to cover daytime conditions. We manually annotated a subset of the images using the following classes: 1) pedestrian (904), 2) dog walker (180), 3) bicyclist (253), or 4) empty (904), i.e., a skewed dataset with 8\% dog walkers and 11\% bicyclists.

The previous study had a two-fold goal. First, we explored neural network pruning by reducing the size of the well-known VGG16 architecture~\cite{VGG14}. Second, we demonstrated how a CycleGAN~\cite{Zhu_2017_ICCV} could be used to extend the ODD to nighttime conditions. In this paper, we experiment with Grad-CAM on one of the pruned DNNs from our previous work, i.e., a DNN with roughly 3.3 million trainable parameters and four convolutional layers. Pre-trained weights for the network are available on GitHub.

\section{Proof-of-Concept} \label{sec:poc}
Our work is based on the open Grad-CAM implementation by Chollet~\cite{fchollet}. However, we extend the work in two ways. First, we provide a more flexible solution that gives test engineers a simpler way to test a CNN by parameterizing the model, its class labels, and the test dataset. We believe the implementation could be easily integrated in our future MLOps pipeline. Second, we extend the original Grad-CAM, which only takes one image as input and provides a Grad-CAM for that specific image, with the concept of a combined Grad-CAM for the average activation of the entire dataset. Hopefully, the combined Grad-CAM will add new insights into how a CNN works on a dataset. Our implementation is designed to flexibly support CNN architectures with a final convolutional layer. All source code is available on GitHub under an MIT license\footnote{https://github.com/augustlidfeldt/ETSN20}. We tested the code on our own CNNs, as well as three pre-trained standard models, i.e., VGG16, VGG19, ResNet50.

\begin{figure}
    \centering
    \includegraphics[width=230pt]{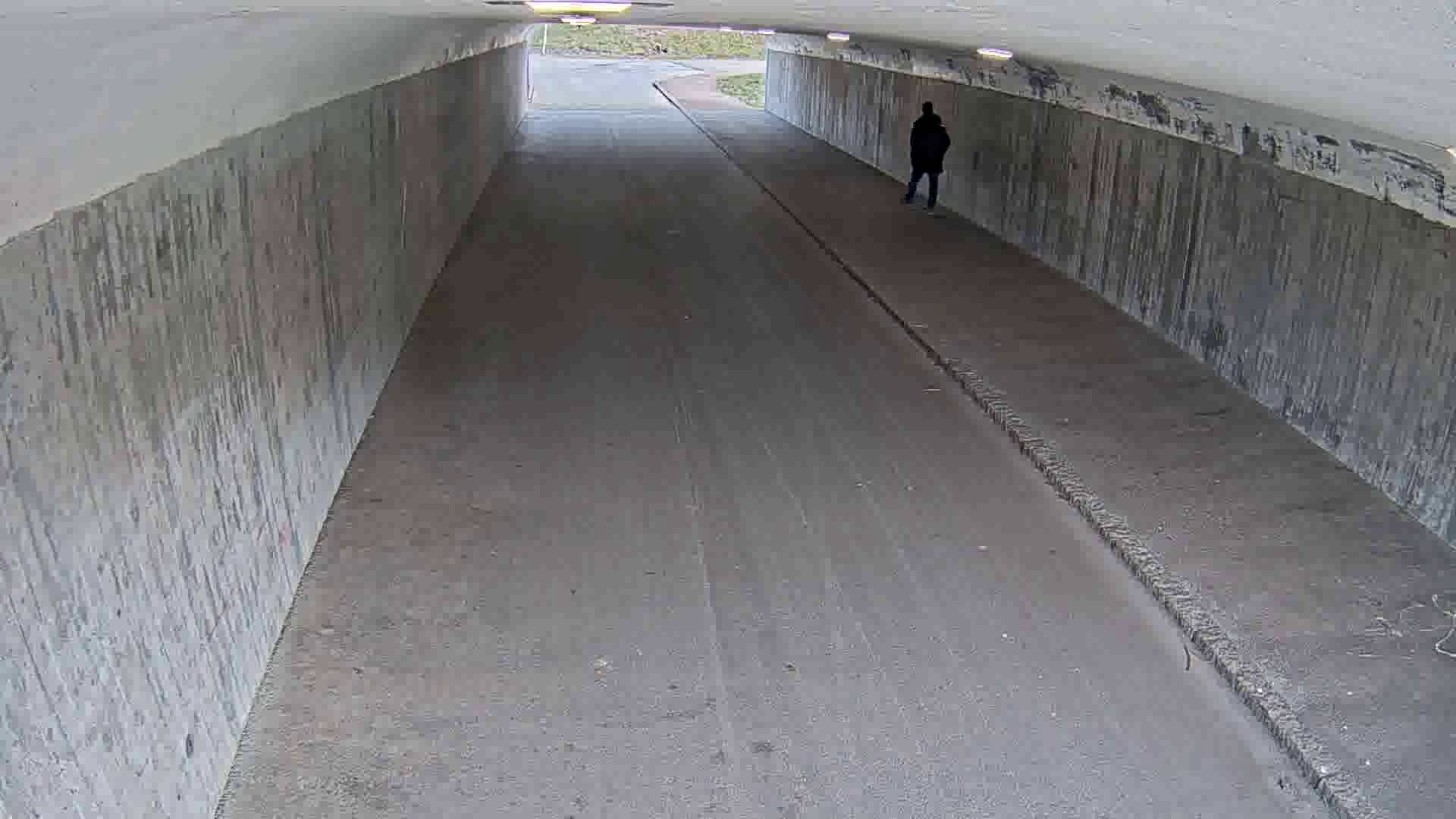}
    \caption{Example pedestrian in the daytime dataset (within the ODD).}
    \label{fig:ped_day}
\end{figure}

\begin{figure}
    \centering
    \includegraphics[width=230pt]{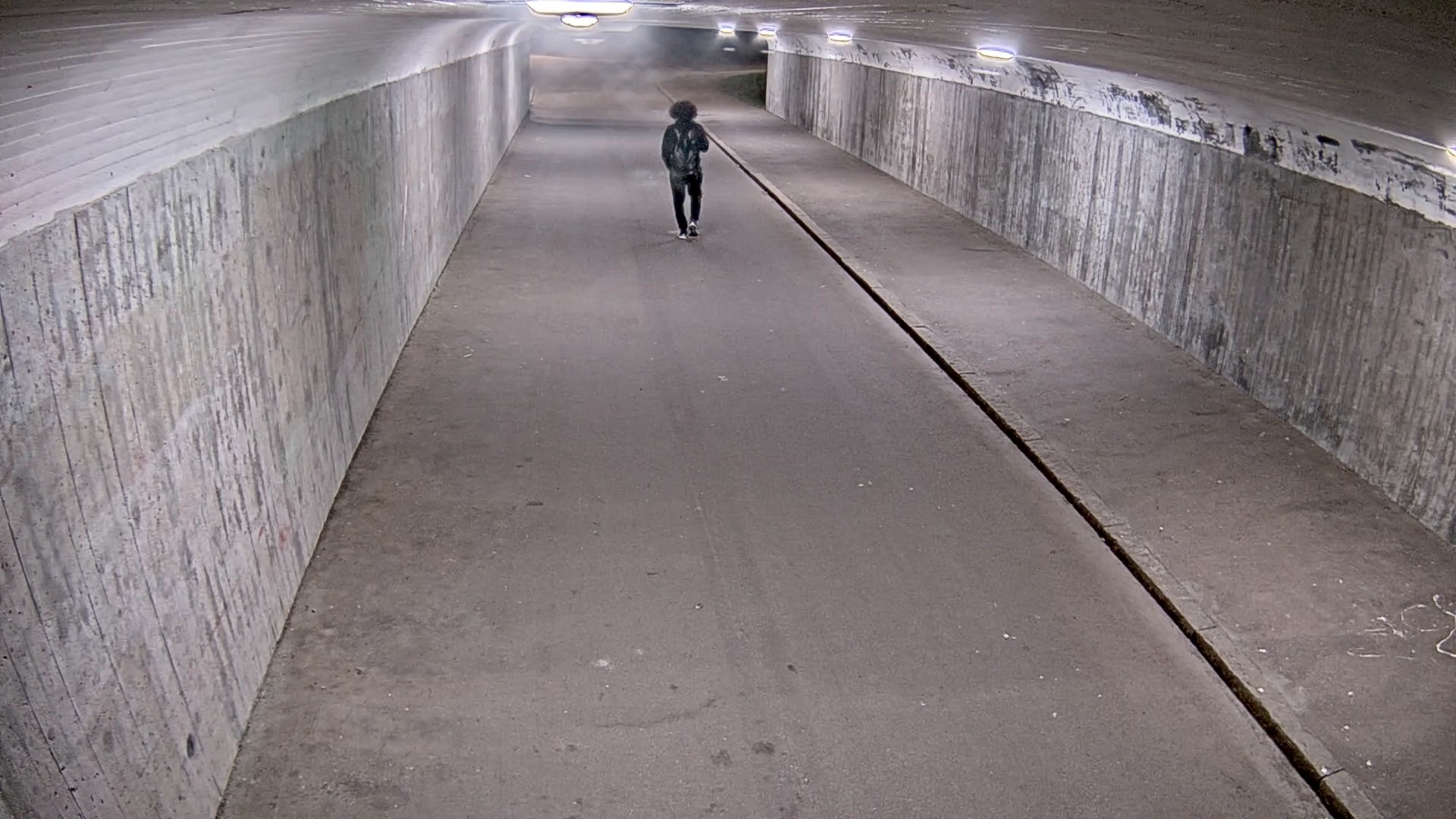}
    \caption{Example pedestrian in the nighttime dataset (outside of the ODD).}
    \label{fig:ped_night}
\end{figure}

We present results from using Grad-CAM on an image recognition model from the Helsingborg underpass described in Section~\ref{sec:case}. All images have an original size of 1920×1080 pixels, but were resized to 224×224 pixels. The daytime test dataset contains 110 images containing pedestrians, bicyclists, and dog walkers under normal operating conditions, i.e., within the daytime ODD. The model under study achieves an 80\% classification accuracy on the daytime dataset. For comparison, we also explore a small test dataset with pictures that are outside the ODD. We already know that these are often misclassified~\cite{lidfeldt2020enabling} -- now we will use Grad-CAM to investigate why this happens. Figures~\ref{fig:ped_day} and~\ref{fig:ped_night} show example images from the daytime and nighttime datasets, respectively. 

Next, we present a selection of Grad-CAM examples. Each Grad-CAM is visualized as a heatmap superimposed over the original image. The heatmaps use a sequential palette with multiple color hues, from colder blue to warmer red, to graphically display relative neuron activation for each pixel. Very low activations are presented with a blueish shade, whereas the red color depict pixels that trigger high activations. Yellow areas in the images display moderate activations. In each figure caption, we report the confidence level of the classification model for that particular prediction.

\begin{figure}
    \centering
    \includegraphics[width=230pt]{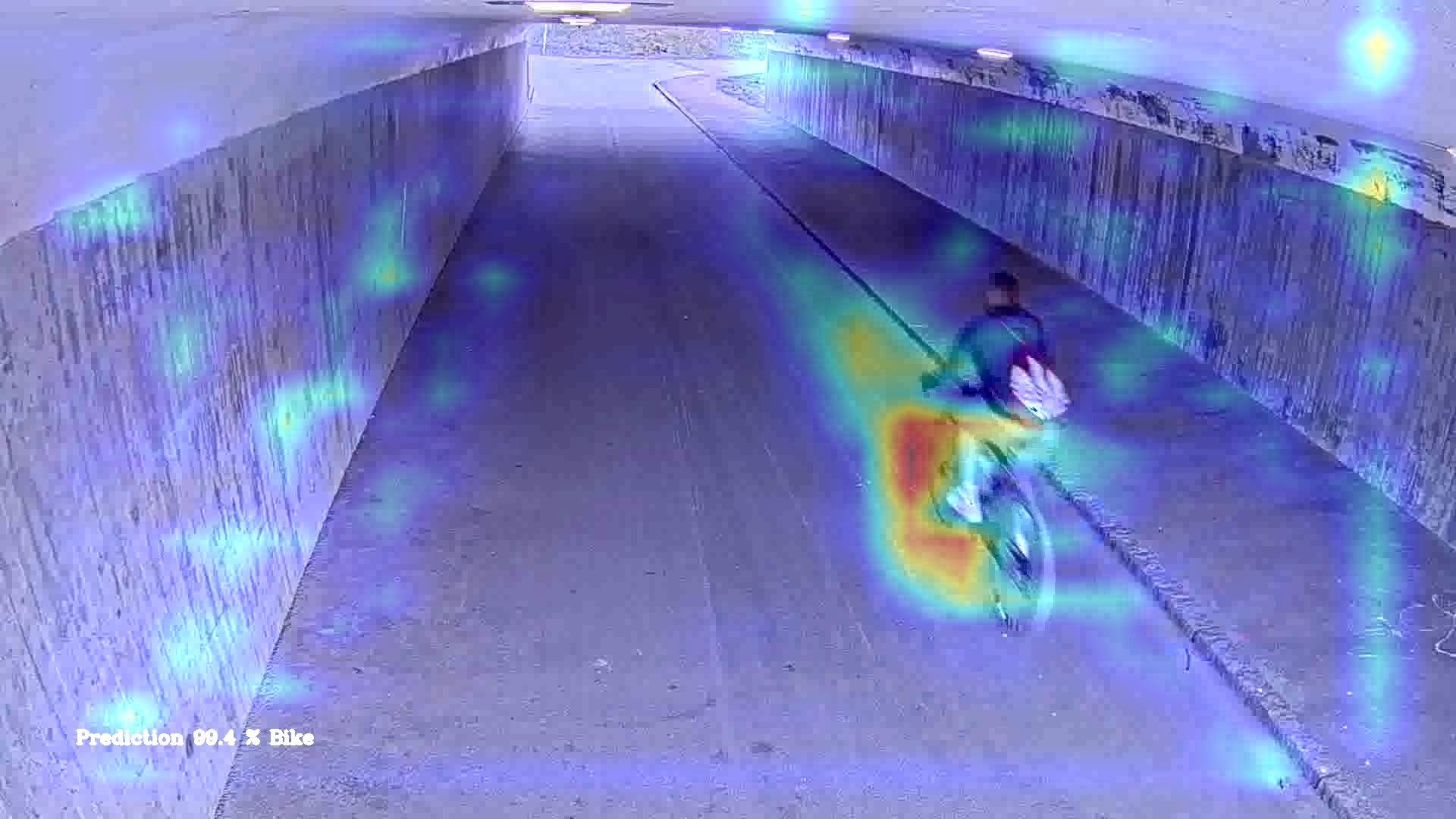}
    \caption{Correctly classified bicyclist (Conf=99.4\%).}
    \label{fig:bike_day_correct}
\end{figure}

\begin{figure}
    \centering
    \includegraphics[width=230pt]{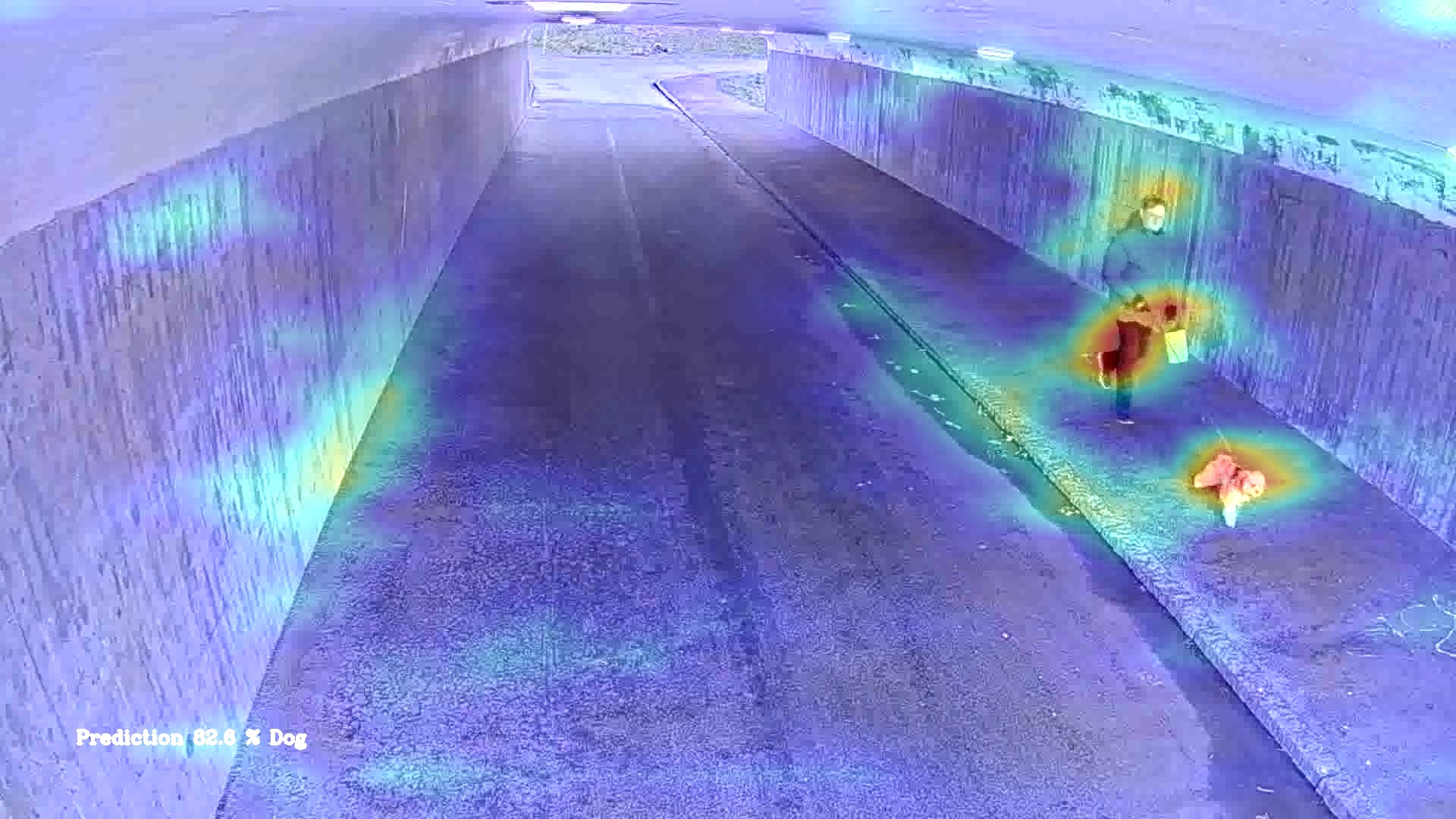}
    \caption{Correctly classified dog walker (Conf=82.6\%).}
    \label{fig:dog_day_correct}
\end{figure}

We first show individual Grad-CAMs, i.e., visual explanations of classifications of single images. Figures~\ref{fig:bike_day_correct} and~\ref{fig:dog_day_correct} show examples of correct classifications of a bicyclist and a dog walker, respectively. The image of the dog walker activate neuron activations corresponding to pixels on the human body, especially the legs and the head, and the small white dog. For the bicyclist, the highest activations are around the front wheel. Moreover, the upper leg and the bicycle saddle are activated, potentially indicating that the model identified the seated position of the human body as a distinguishing feature. On the other hand, both Grad-CAMs denote activated pixels far from the relevant object, e.g., the top right corner and the opposing wall. Such activations might suggest the presence of unwanted bias.  

\begin{figure}
    \centering
    \includegraphics[width=230pt]{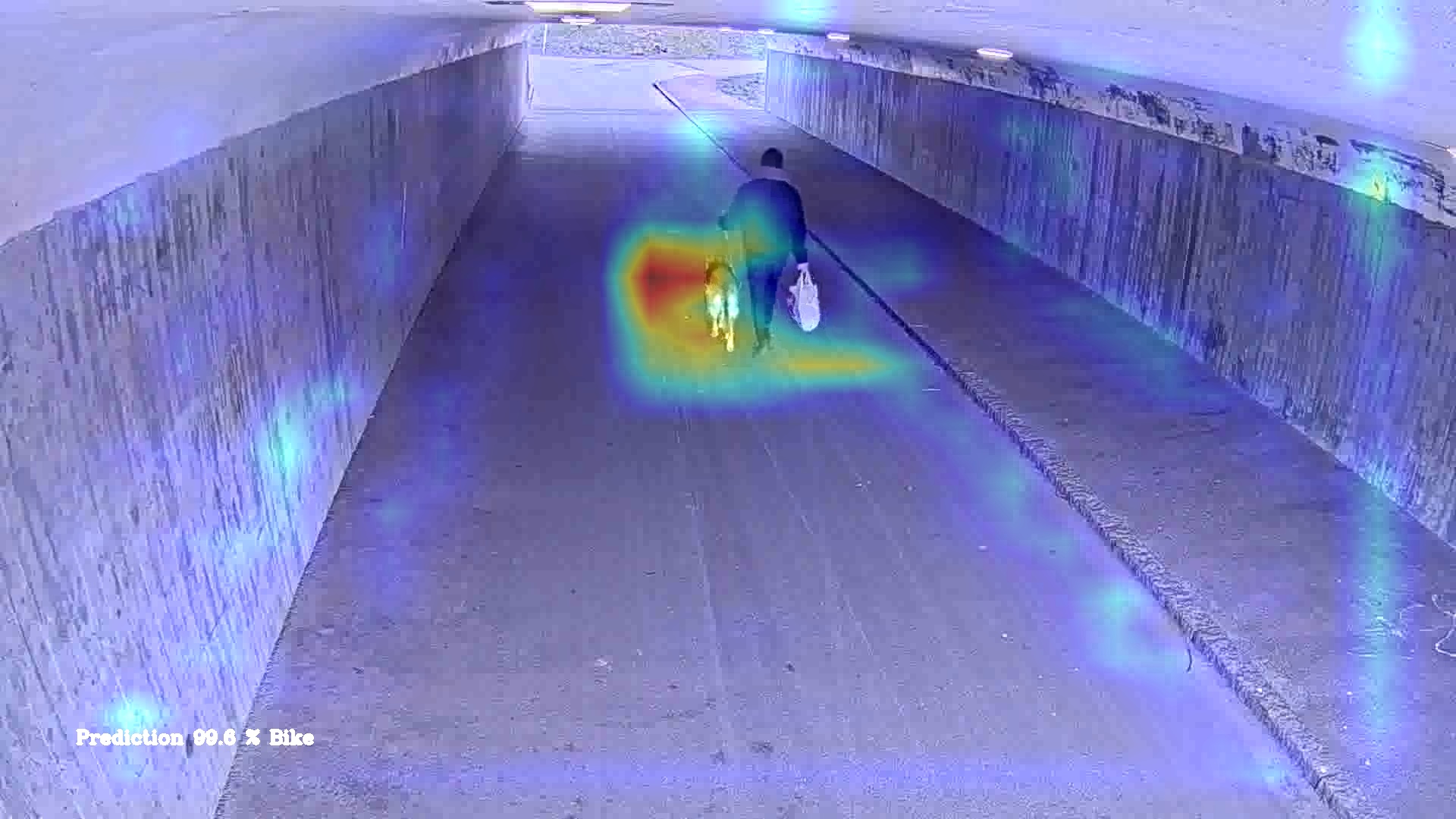}
    \caption{Classification of daytime dog walker as bicyclist (Conf=99.6\%).}
    \label{fig:dog_day_incorrect}
\end{figure}

Figure~\ref{fig:dog_day_incorrect} shows an example of a Grad-CAM for a missclassification. The model is activated around the dog, especially on its left side. The model certainly localizes the area of interest in the image, but incorrectly predicts the bicyclist label. For this particular image, it is possible that the location of dog walker was the distinguishing feature. Most dog walkers in the training set appear on the sidewalk, not on the bike lane. 

\begin{figure}
    \centering
    \includegraphics[width=230pt]{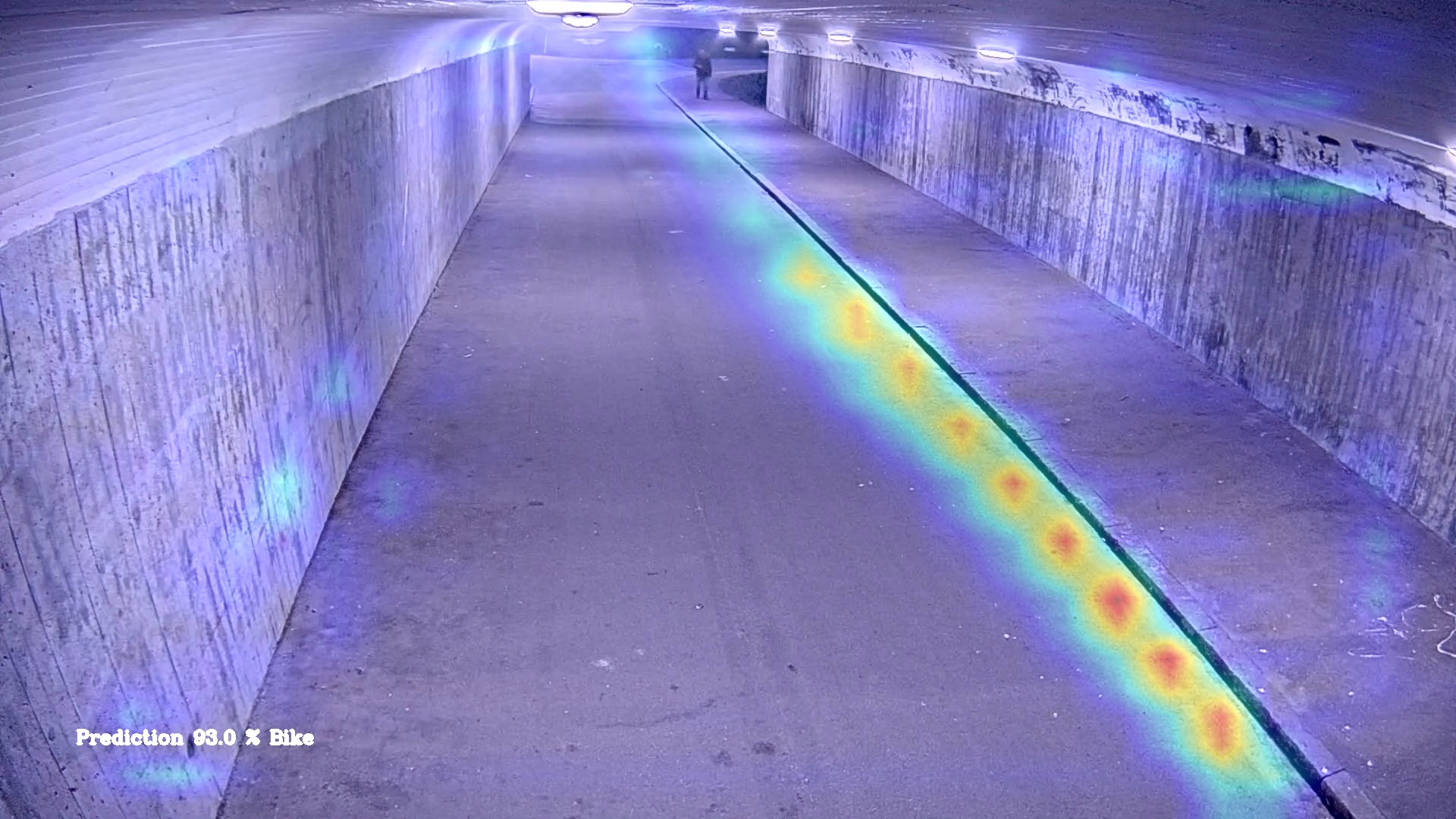}
    \caption{Classification of nighttime pedestrian as bicyclist (Conf=93.0\%).}
    \label{fig:ped_night_incorrect}
\end{figure}

\begin{figure}
    \centering
    \includegraphics[width=230pt]{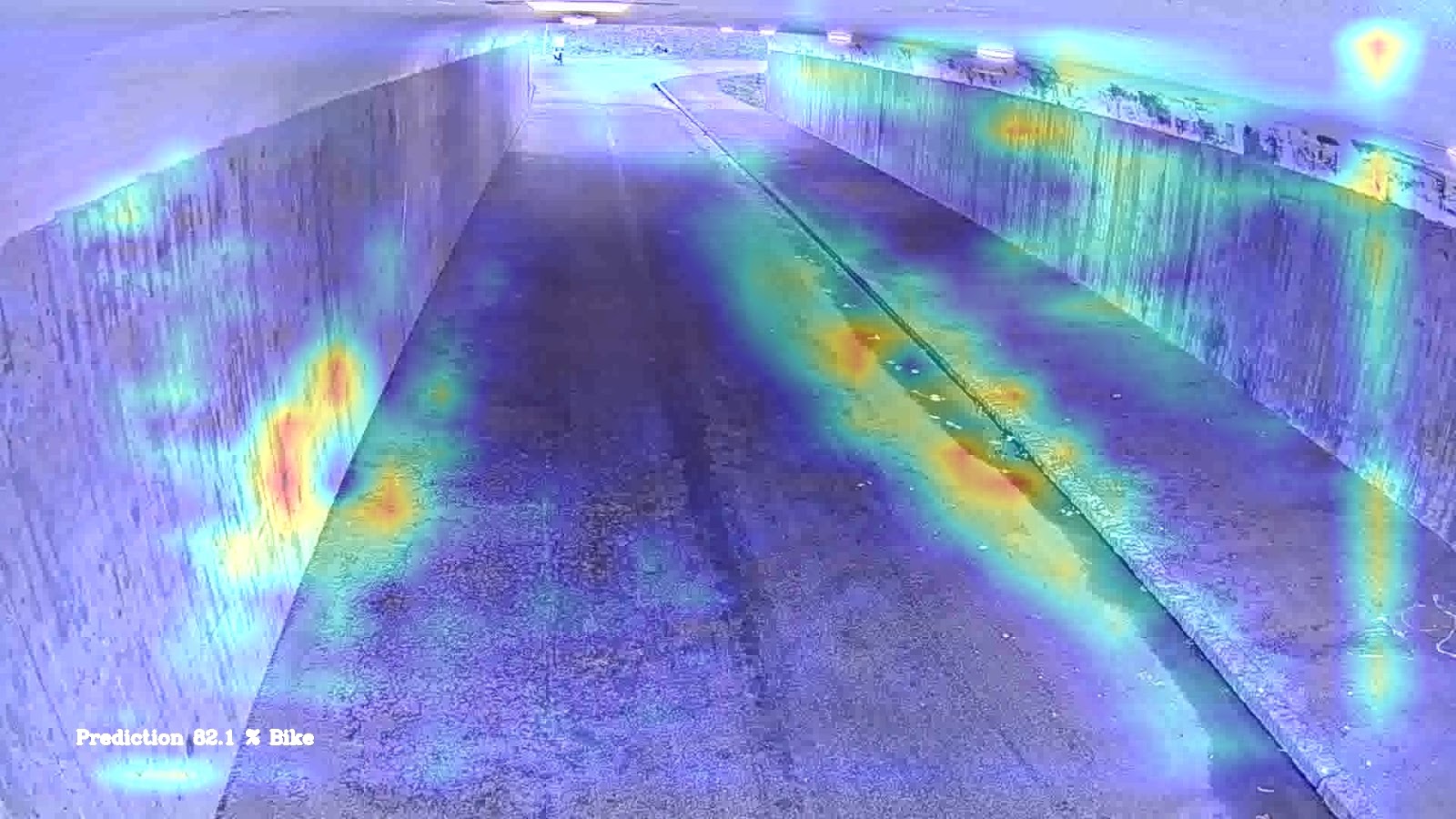}
    \caption{Example of a bicylist example in the training set. The bicyclist is very far away and there is a dark water puddle along the curb.}
    \label{fig:waterpuddle}
\end{figure}

Figure~\ref{fig:ped_night_incorrect} displays an example of a missclassification of a nighttime image, i.e., outside of the ODD. A pedestrian is visible in the far end of the underpass, but the model classifies the image as another bicyclist. The heatmap shows that pixels along the entire curb activated the model, suggesting that it has learned a flawed representation of the bicyclist class. Based on this new insight, we manually inspected the images with a bicyclist label in the training set. Note that all these images are from the daytime domain. We found that many of the images contain bikers in the far end of the tunnel. Furthermore, a phenomenon depicted in Figure~\ref{fig:waterpuddle}, several daytime images contain a dark water puddle along the curb. We find it reasonable that the sharper shadows in artificially illuminated nighttime underpass resulted in the model finding imaginary water, thus predicting the bicyclist label -- a clear illustration of biased training data.

\begin{figure}
    \centering
    \includegraphics[width=230pt]{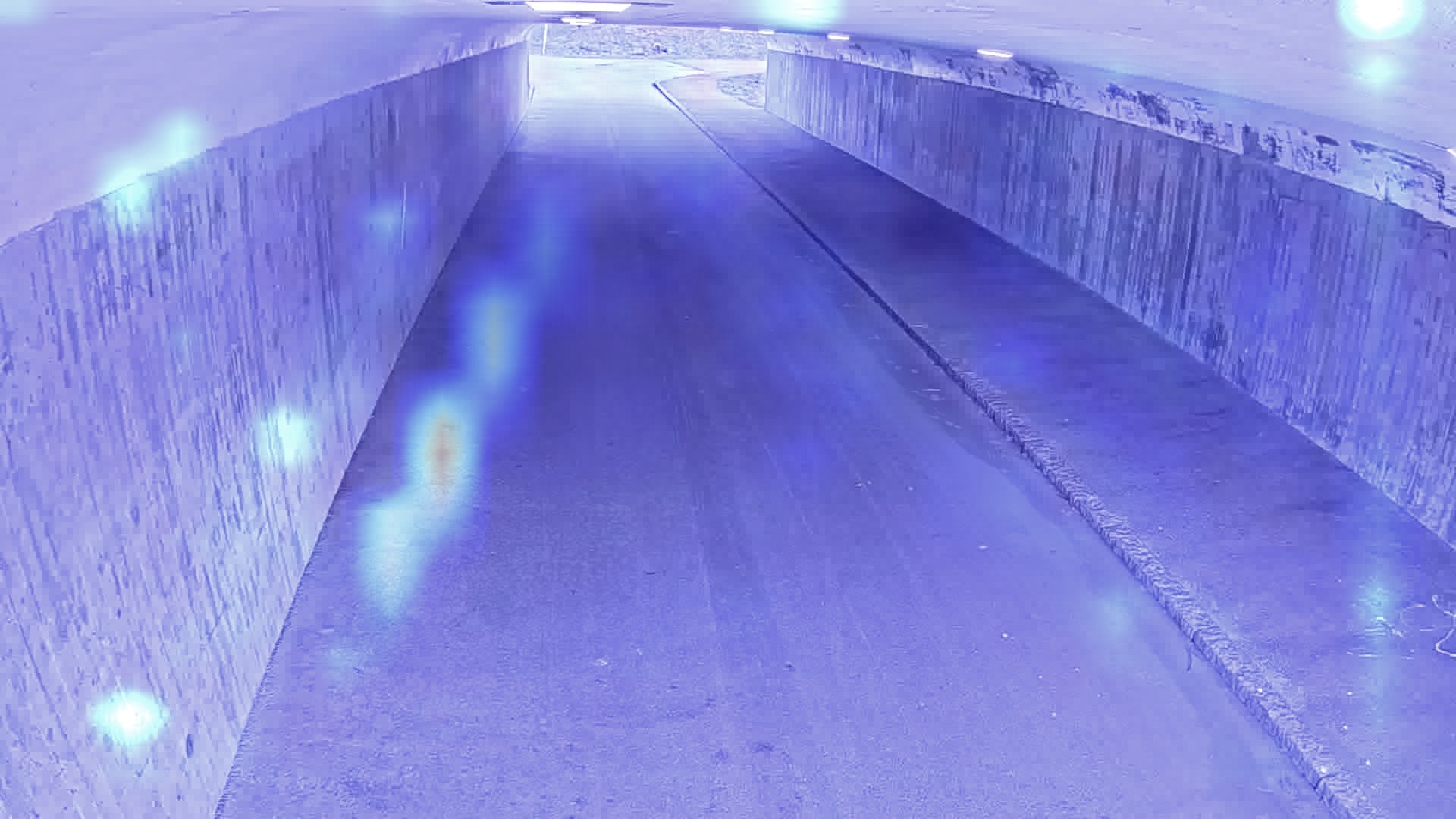}
    \caption{A combined Grad-CAM showing average activations for the test set.}
    \label{fig:combined}
\end{figure}

Finally, we showcase how a combined Grad-CAM can provide an additional perspective. Figure~\ref{fig:combined} shows a heatmap visually explaining the average activations for the entire test set. The heatmap shows relatively high activations on the left side of the bike lane. While this could be valid, it is more suspicious to find high activations in the top right corner and high up on the left side wall. We manually examined the training set but did not find anything in these regions that should affect the model when classifying whether an image contains an empty tunnel, a pedestrian, a bicyclist or a dog walker -- another indication of bias in the dataset.

\section{Discussion and Future Work} \label{sec:disc}
We posit that integrating automated Grad-CAM generation and evaluation as a step in an MLOps pipeline can help development organizations working with CNN-based perception systems. We revisit the research questions and argue that organizations can use Grad-CAM in two primary ways. First, \textit{the visual explanations added by Grad-CAM heatmaps can support organizations when demonstrating that their ML-based systems are ALTAI compliant} (RQ1). Second, we envision that \textit{automated analysis of Grad-CAM heatmaps can be used as part of test automation to find indications of biased ML models} (RQ2).

Regarding \textit{ALTAI compliance}, we argue that five of the key requirements of trustworthy AI would be supported by the explainability provided by Grad-CAM. 

\begin{itemize}
\item[R1] \textbf{Human agency and oversight}. R1 mandates that oversight mechanisms must be implemented to allow humans to inspect the decisions made. The heatmaps presented in Section~\ref{sec:poc} could constitute an important component in such a mechanism, i.e., enabling \textit{post hoc} analysis.
\item[R2] \textbf{Technical robustness and safety}. In critical applications, it is not enough to merely rely on standard accuracy metrics. Providing developers with tools to understand why a model performs inaccurately, would be a welcome addition when evolving a system to reach robustness targets. ML development is highly iterative, and Grad-CAM heatmaps could be a tool to detect deteriorating accuracy and distributional shifts.
\item[R4] \textbf{Transparency}. Grad-CAM is an initiative related to explainable AI, thus adding corresponding heatmaps would support R4. Users and developers could analyze why certain classifications turn out incorrect, thus providing insights into inherent capabilities and limitations.
\item[R5] \textbf{Diversity, non-discrimination and fairness}. Grad-CAM heatmaps could be used to create evidence that a classification model activates based on the same type of pixels for different user subgroups. In Section~\ref{sec:poc}, we primarily illustrated bias towards non-sensitive features such as water puddles and physical infrastructure. However, Grad-CAM could also be used to investigate protected attributes~\cite{vogelsang2019requirements}, e.g., clothing, ethnicity, or disabilities.
\item[R7] \textbf{Accountability}. R7 requires implementation of mechanisms to enable auditability, including the assessment of algorithms and the underlying training data. Grad-CAM heatmaps could help external auditors, also non-experts, to interpret what features in an image matters the most to a classification outcome.
\end{itemize}

To ensure privacy we only used non-sensitive images that do not reveal personal information. However, there is a need for further studies to better understand if explainable AI could supports R3, \textit{Privacy and data governance}. Furthermore, if increased explainability could support R6, \textit{Societal and environmental well-being} also requires further studies.

\begin{figure}
    \centering
    \includegraphics[width=230pt]{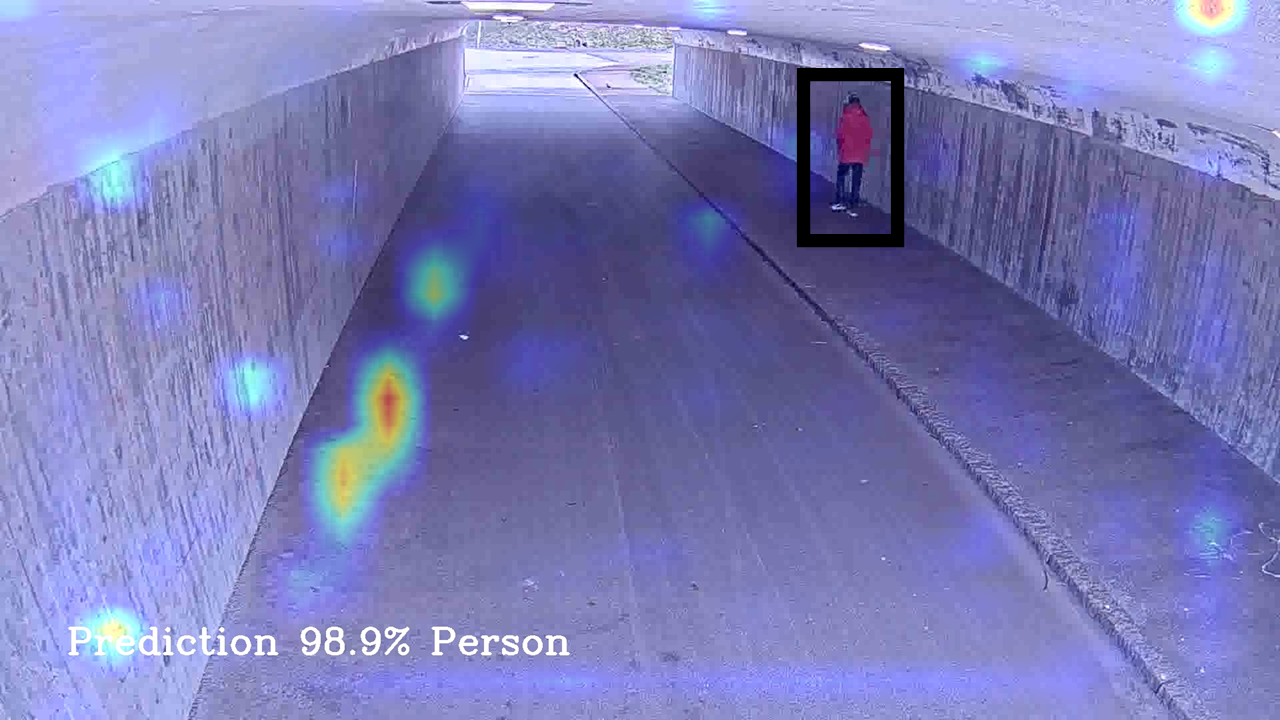}
    \caption{Correctly classified pedestrian (Conf=98.9\%). However, the missing overlap between the heatmap and the bounding box should trigger a failed test case.}
    \label{fig:boundingbox}
\end{figure}

Regarding \textit{automated Grad-CAM analysis}, we believe that heatmaps could be successfully used to compare against test images annotated with ground truth bounding boxes. Figure~\ref{fig:boundingbox} shows a correctly classified pedestrian. However, the pixels that activated the model are not within the black bounding box around the human. Used in an automated test suite, this should cause the test case to fail. The size of the bounding box could be used to specify the tolerance of the test case. For the most precise models, pixel-wise segmentation would be an option. Our current work is limited to image recognition models, but we plan to extend our work to support object detection models such as YOLO and RCNN~\cite{kim2020comparison}.

There are several limitations with the initial results that we present in this paper. First, we present results based on a CNN that has been trained with a very small dataset. While we found the generated Grad-CAM heatmaps highly useful in guiding future efforts in data collection to increase the training set, there is a risk that CNNs adequately trained on large datasets would result in less interpretable heatmaps. However, our initial experience from using heatmaps on common CNN architectures with pretrained weights suggests no problems, neither does the original work by Selvaraju \textit{et al.}~\cite{selvaraju2017grad-cam}. Second, we believe that the feasibility of using Grad-CAM heatmaps generalizes from our example with a static camera in a pedestrian underpass to a critical application that relies on a forward-facing camera mounted on a vehicle. This is obviously a big step that must be tackled by incremental research. Our immediate next step will be to conduct a similar proof-of-concept in the automotive domain, with a focus on SOTIF rather than ALTAI compliance.

\section{Conclusion} \label{sec:conc}
Machine learning is an essential component in the perception systems used in cyber-physical systems. Unfortunately, the inner workings of trained deep neural networks are hard to understand for human analysts. Limited explainability is a major obstacle to tackle before such systems can be certified for use in critical applications. Consequently, the area of explainable AI is rapidly evolving. One promising technique originating in this line of research is Gradient-weighted Class Activation Mapping (Grad-CAM)~\cite{selvaraju2017grad-cam}.

We presented a proof-of-concept of how we used Grad-CAM heatmaps to explain which pixels were decisive for a trained neural network classifier. Furthermore, we discuss how increased explainability through Grad-CAM supports several of the key requirements for Trustworthy AI as defined by an expert group established by the the European Commission. For our particular classifier, trained on image data from a camera mounted in an underpass, we identify several indications of bias in the training data. We use Grad-CAM both for individual images and the test set's average activations -- providing actionable pointers on how to guide future data collection.

Software and systems engineering is increasingly relying on continuous engineering approaches. In this paper, we discussed how we plan to add analysis of Grad-CAM heatmaps as a pipe segment in an automated MLOps pipeline. We present an initial version of this block in this paper, available on GitHub under an MIT licence. As future work, we plan to extend the approach from testing object recognition in the static underpass installation to testing object detection for forward-facing cameras used in advanced driver-assistance systems. Our first step in this direction will be to integrate this approach to testing in simulated environments for our ongoing work on an automated pedestrian emergency braking system.

\section*{Acknowledgements}
This work was funded by Kompetensfonden at Campus Helsingborg, Lund University, Sweden. Furthermore, the project received financial support from the SMILE~III project financed by Vinnova, FFI, Fordonsstrategisk forskning och innovation under the grant number: 2019-05871 and the ECSEL Joint Undertaking (JU) under grant agreement No 876852 (VALU3S).

\bibliographystyle{IEEEtran}
\bibliography{bibliography}

% Generated by IEEEtran.bst, version: 1.13 (2008/09/30)
\begin{thebibliography}{10}
\providecommand{\url}[1]{#1}
\csname url@samestyle\endcsname
\providecommand{\newblock}{\relax}
\providecommand{\bibinfo}[2]{#2}
\providecommand{\BIBentrySTDinterwordspacing}{\spaceskip=0pt\relax}
\providecommand{\BIBentryALTinterwordstretchfactor}{4}
\providecommand{\BIBentryALTinterwordspacing}{\spaceskip=\fontdimen2\font plus
\BIBentryALTinterwordstretchfactor\fontdimen3\font minus
  \fontdimen4\font\relax}
\providecommand{\BIBforeignlanguage}[2]{{%
\expandafter\ifx\csname l@#1\endcsname\relax
\typeout{** WARNING: IEEEtran.bst: No hyphenation pattern has been}%
\typeout{** loaded for the language `#1'. Using the pattern for}%
\typeout{** the default language instead.}%
\else
\language=\csname l@#1\endcsname
\fi
#2}}
\providecommand{\BIBdecl}{\relax}
\BIBdecl

\bibitem{falcini_deep_2017}
F.~Falcini, G.~Lami, and A.~Costanza, ``Deep {Learning} in {Automotive}
  {Software},'' \emph{IEEE Software}, vol.~34, no.~3, pp. 56--63, 2017.

\bibitem{high-level_expert_group_on_artificial_intelligence_ethics_2019}
{High-Level Expert Group on Artificial Intelligence}, ``Ethics {Guidelines} for
  {Trustworthy} {Artificial} {Intelligence},'' European Commission, Brussels,
  Belgium, Tech. Rep., 2019.

\bibitem{adadi_peeking_2018}
A.~Adadi and M.~Berrada, ``Peeking {Inside} the {Black}-{Box}: {A} {Survey} on
  {Explainable} {Artificial} {Intelligence} ({XAI}),'' \emph{IEEE Access},
  vol.~6, pp. 52\,138--52\,160, 2018.

\bibitem{shen2020interpretability}
O.~Shen, ``Interpretability in {ML}: A broad overview,'' \emph{The Gradient},
  2020.

\bibitem{azulay_why_2019}
A.~Azulay and Y.~Weiss, ``\BIBforeignlanguage{en}{Why {Do} {Deep}
  {Convolutional} {Networks} {Generalize} so {Poorly} to {Small} {Image}
  {Transformations}?}'' \emph{\BIBforeignlanguage{en}{Journal of Machine
  Learning Research}}, vol.~20, p.~25, 2019.

\bibitem{selvaraju2017grad-cam}
R.~R. {Selvaraju}, M.~{Cogswell}, A.~{Das}, R.~{Vedantam}, D.~{Parikh}, and
  D.~{Batra}, ``Grad-cam: Visual explanations from deep networks via
  gradient-based localization,'' in \emph{Proc. of the IEEE International
  Conference on Computer Vision}, 2017, pp. 618--626.

\bibitem{Trokielewicz2018}
M.~{Trokielewicz}, A.~{Czajka}, and P.~{Maciejewicz}, ``Presentation attack
  detection for cadaver iris,'' in \emph{Proc. of the 9th International
  Conference on Biometrics Theory, Applications and Systems}, 2018, pp. 1--10.

\bibitem{gorski2020towards}
L.~Gorski, S.~Ramakrishna, and J.~M. Nowosielski, ``Towards grad-cam based
  explainability in a legal text processing pipeline,'' \emph{arXiv preprint
  arXiv:2012.09603}, 2020.

\bibitem{9307643}
G.~{Iadarola}, F.~{Martinelli}, F.~{Mercaldo}, and A.~{Santone}, ``Evaluating
  deep learning classification reliability in android malware family
  detection,'' in \emph{Proc. of the IEEE International Symposium on Software
  Reliability Engineering Workshops}, 2020, pp. 255--260.

\bibitem{borg_aiq_2021}
M.~Borg, ``The {AIQ} {Meta}-{Testbed}: {Pragmatically} {Bridging} {Academic}
  {AI} {Testing} and {Industrial} {Q} {Needs},'' in \emph{Software {Quality}:
  {Future} {Perspectives} on {Software} {Engineering} {Quality}}, D.~Winkler,
  S.~Biffl, D.~Mendez, M.~Wimmer, and J.~Bergsmann, Eds.\hskip 1em plus 0.5em
  minus 0.4em\relax Cham: Springer International Publishing, 2021, pp. 66--77.

\bibitem{lidfeldt2020enabling}
A.~Lidfeldt, D.~Isaksson, L.~Hedlund, S.~{\AA}berg, M.~Borg, and E.~Larsson,
  ``Enabling image recognition on constrained devices using neural network
  pruning and a cyclegan,'' in \emph{Proc. of the 10th International Conference
  on the Internet of Things Companion}, 2020, pp. 1--14.

\bibitem{marijan2019challenges}
D.~Marijan, A.~Gotlieb, and M.~K. Ahuja, ``Challenges of testing machine
  learning based systems,'' in \emph{2019 IEEE International Conference On
  Artificial Intelligence Testing (AITest)}.\hskip 1em plus 0.5em minus
  0.4em\relax IEEE, 2019, pp. 101--102.

\bibitem{salay2018using}
R.~Salay and K.~Czarnecki, ``Using machine learning safely in automotive
  software: An assessment and adaption of software process requirements in
  {ISO} 26262,'' \emph{arXiv preprint arXiv:1808.01614}, 2018.

\bibitem{borg2019safely}
M.~Borg, C.~Englund, K.~Wnuk, B.~Durann, C.~Lewandowski, S.~Gao, Y.~Tan,
  H.~Kaijser, H.~L{\"o}nn, and J.~T{\"o}rnqvist, ``Safely entering the deep: A
  review of verification and validation for machine learning and a challenge
  elicitation in the automotive industry,'' \emph{Journal of Automotive
  Software Engineering}, vol.~1, no.~1, pp. 1--19, 2019.

\bibitem{international_organization_for_standardization_isopas_2019}
{International Organization for Standardization}, ``{ISO}/{PAS} 21448:2019
  {Road} {Vehicles} - {Safety} of the {Intended} {Functionality},'' Tech. Rep.
  ISO/PAS 21448, 2019.

\bibitem{hao2019training}
K.~Hao, ``Training a single {AI} model can emit as much carbon as five cars in
  their lifetimes,'' \emph{MIT Technology Review}, 2019.

\bibitem{tesla}
\BIBentryALTinterwordspacing
Tesla, ``Tesla autopilot.'' [Online]. Available:
  \url{https://www.tesla.com/autopilotAI}
\BIBentrySTDinterwordspacing

\bibitem{zhang_machine_2020}
J.~M. Zhang, M.~Harman, L.~Ma, and Y.~Liu, ``Machine {Learning} {Testing}:
  {Survey}, {Landscapes} and {Horizons},'' \emph{IEEE Transactions on Software
  Engineering}, pp. 1--1, 2020.

\bibitem{pei_deepxplore_2019}
K.~Pei, Y.~Cao, J.~Yang, and S.~Jana, ``{DeepXplore}: {Automated} {Whitebox}
  {Testing} of {Deep} {Learning} {Systems},'' \emph{Communications of the ACM},
  vol.~62, no.~11, pp. 137--145, 2019.

\bibitem{kim_guiding_2019}
J.~Kim, R.~Feldt, and S.~Yoo, ``Guiding {Deep} {Learning} {System} {Testing}
  {Using} {Surprise} {Adequacy},'' in \emph{Proc. of the 41st {International}
  {Conference} on {Software} {Engineering}}, May 2019, pp. 1039--1049.

\bibitem{patel2019adaptive}
N.~Patel, P.~Krishnamurthy, S.~Garg, and F.~Khorrami, ``Adaptive adversarial
  videos on roadside billboards: Dynamically modifying trajectories of
  autonomous vehicles,'' in \emph{Proc. of the IEEE/RSJ International
  Conference on Intelligent Robots and Systems}, 2019.

\bibitem{high-level_expert_group_on_artificial_intelligence_assessment_2020}
{High-Level Expert Group on Artificial Intelligence}, ``Assessment {List} for
  {Trustworthy} {Artificial} {Intelligence},'' European Commission, Brussels,
  Belgium, Tech. Rep., 2020.

\bibitem{gyllenhammar2020towards}
M.~Gyllenhammar, R.~Johansson, F.~Warg, D.~Chen, H.-M. Heyn, M.~Sanfridson,
  J.~S{\"o}derberg, A.~Thors{\'e}n, and S.~Ursing, ``Towards an operational
  design domain that supports the safety argumentation of an automated driving
  system,'' in \emph{Proc. of the 10th European Congress on Embedded Real Time
  Systems}, 2020.

\bibitem{VGG14}
K.~Simonyan and A.~Zisserman, ``Very deep convolutional networks for
  large-scale image recognition,'' 2014.

\bibitem{Zhu_2017_ICCV}
J.-Y. Zhu, T.~Park, P.~Isola, and A.~A. Efros, ``Unpaired image-to-image
  translation using cycle-consistent adversarial networks,'' in \emph{Proc. of
  the 2017 IEEE International Conference on Computer Vision}, Oct 2017.

\bibitem{fchollet}
\BIBentryALTinterwordspacing
F.~Chollet, ``5.4-visualizing-what-convnets-learn,'' 2017. [Online]. Available:
  \url{\url{https://github.com/fchollet/deep-learning-with-python-notebooks/blob/master/5.4-visualizing-what-convnets-learn.ipynb}}
\BIBentrySTDinterwordspacing

\bibitem{vogelsang2019requirements}
A.~Vogelsang and M.~Borg, ``Requirements engineering for machine learning:
  Perspectives from data scientists,'' in \emph{Proc. of the IEEE 27th
  International Requirements Engineering Conference Workshops}.\hskip 1em plus
  0.5em minus 0.4em\relax IEEE, 2019, pp. 245--251.

\bibitem{kim2020comparison}
J.-a. Kim, J.-Y. Sung, and S.-h. Park, ``Comparison of {Faster-RCNN}, {YOLO},
  and {SSD} for real-time vehicle type recognition,'' in \emph{Proc. of the
  IEEE International Conference on Consumer Electronics-Asia}.\hskip 1em plus
  0.5em minus 0.4em\relax IEEE, 2020, pp. 1--4.

\end{thebibliography}
\clearpage
\end{document}